# Common Sense Knowledge, Ontology and Text Mining for Implicit Requirements

Onyeka Emebo[1,2], Aparna S. Varde[1], Olawande Daramola [2]
1. Department of Computer Science, Montclair State University, Montclair, NJ, USA.
2. Department of Computer and Information Sciences, Covenant University, Ota, Nigeria
emeboo@montclair.edu, vardea@montclair.edu, olawande.daramola@covenantuniversity.edu.ng

*Abstract*— **The ability of a system to meet its requirements is a strong determinant of success. Thus effective requirements specification is crucial. Explicit Requirements are well-defined needs for a system to execute. IMplicit Requirements (IMRs) are assumed needs that a system is expected to fulfill though not elicited during requirements gathering. Studies have shown that a major factor in the failure of software systems is the presence of unhandled IMRs. Since relevance of IMRs is important for efficient system functionality, there are methods developed to aid the identification and management of IMRs. In this paper, we emphasize that Common Sense Knowledge, in the field of Knowledge Representation in AI, would be useful to automatically identify and manage IMRs. This paper is aimed at identifying the sources of IMRs and also proposing an automated support tool for managing IMRs within an organizational context. Since this is found to be a present gap in practice, our work makes a contribution here. We propose a novel approach for identifying and managing IMRs based on combining three core technologies: common sense knowledge, text mining and ontology. We claim that discovery and handling of unknown and non-elicited requirements would reduce risks and costs in software development.**

## I. INTRODUCTION

The challenge of identifying and managing implicit requirements has developed to be a crucial subject in the field of requirements engineering. In [7] it was stated "When critical knowledge, goals, expectations or assumptions of key stakeholders remain hidden or unshared then poor requirements and poor systems are a likely, and costly, consequence." With the relevance of implicit requirements (IMRs) being identified and related to the efficient functionality of any developed system, there have been different proposals as well as practical methodologies developed to aid the identification and management of IMRs. Common Sense Knowledge (CSK) is an area that involves making a computer or another machine understand basic facts as intuitively as a human would. It is an area in the realm of Knowledge Representation (KR) which involves paradigms for adequate depiction of knowledge in Artificial Intelligence (AI). The area of CSK is being researched for its use in identification and capturing of implicit requirements.

Since AI is aimed at enabling machines solve problems like humans, there is a need for common sense knowledge in AI systems to enhance problem-solving. This not only involves storing what most people know but also the application of that knowledge [8]. In software engineering, the development of systems must meet the needs of the intended user. However the very fact that CSK is common, not all knowledge and requirements that entail common sense, will be captured or expressed by the expected user. As Polanyi describes "We know more than we can tell". It is therefore the responsibility of the software developer to capture as well as manage the unexpressed requirements in the development of a suitable and satisfactory system. The application of Common Sense Knowledge can improve the identification as well as management of IMRs. Common Sense Knowledge CSK) is defined in [3] as a collection of simple facts about people and everyday life, such as "Things fall down, not up", and "People eat breakfast in the morning". In [7], the authors describe CSK as a tremendous amount and variety of knowledge of default assumptions about the world, which is shared by (possibly a group of) people and seems so fundamental and obvious that it usually does not explicitly appear in people's communications. CSK is mainly characterized by its implicitness.

From the literature, it is observed that a number of reasons have caused the emergence of implicit requirements some of which include; i) When a software organization develops a product in a new domain and ii) as a result of knowledge gap between developers and stakeholders due to the existence of implicit knowledge.

Given this background, we claim that CSK will aid in the identification of IMRs useful for domain-specific knowledge bases. This will be useful for storing domain concepts, relations and instances for onward use in domain related processing, knowledge reuse and discovery. Thus we build an automated IMR support tool based on our proposed framework for managing IMRs using common sense knowledge, ontology and text mining.

The rest of this paper is organized as follows: Section II presents core technologies. Section III reviews related work on IMRs. Section IV describes our automated IMR process framework. Section V describes the use and evaluation of the IMR support tool. Section VI gives the conclusions.

## II. BACKGROUND

In this section, an overview of the concepts relevant in CSK, ontology, text mining and natural language processing is presented. This is useful in order to understand our proposed IMR framework later.

### A. Common Sense Knowledge

Common Sense Knowledge (CSK) according to [3] is a tremendous amount and variety of knowledge of default assumptions about the world, which is shared by people and seems so obvious that it usually does not explicitly appear in

communications. Some characteristics of CSK as identified in the literature are as follows:
- *Share*: A group of people possess and share CSK.
- *Fundamentality*: People have a good understanding of CSK that they tend to take CSK for granted.
- *Implicitness*: People more often don't mention or document CSK explicitly since others also know it.
- *Large-Scale*: CSK is highly diversified and in large quantity.
- *Open-Domain*: CSK is broad in nature covering all aspects of our daily life rather than a specific domain.
- *Default*: CSK are default assumptions about typical cases in everyday life, so most of them might not always be correct.

Previous work on common sense knowledge includes the seminal projects Cyc [9] and WordNet [5], ConceptNet [20], Webchild [31] and the work by [14] and [24]. Cyc has compiled complex assertions such as every human has exactly one father and exactly one mother. WordNet has manually organized nouns and adjectives into lexical classes, with careful distinction between words and word senses. ConceptNet is probably the largest repository of common sense assertions about the world, covering relations such as hasProperty, usedFor, madeOf, motivatedByGoal, etc. Tandon et al. [14] automatically compiled millions of triples of the form <noun relation adjective> by mining n-gram corpora. Lebani & Pianta [24] proposed encoding additional lexical relations for commonsense knowledge into WordNet. WebChild contains triples that connect nouns with adjectives via fine-grained relations like hasShape, hasTaste, evokesEmotion, etc.

### B. Ontology

The term ontology has different meanings. Ontology made and entrance in the field of computer science in the 1990s in association with Knowledge Acquisition. Different definitions have been given to the term "ontology". A basic definition of ontology was given in [37] as an explicit specification of a conceptualization". Author [39] explains it as a special kind of information object or computational artifact while [38] defined an ontology as a "formal specification of a shared conceptualization. Both definitions were merged by [12] hence defining an ontology as a "formal explicit specification of shared conceptualization". Ontologies provide a formal representation of knowledge and the relationships between concepts of a domain. They are used in the requirements specification to guide formal and unambiguous specification of the requirements, particularly in expressing concepts, relations and business rules of domain model with varying degrees of formalization and precision [26].

Formally an Ontology structure $O$ can be defined as [18]

$$O = \{C, R, A^o\}$$

Where:
1. $C$ is a set whose elements are called *concepts*.
2. $R \subseteq C \times C$ is a set whose elements are called *relations*. For $r = (c_1, c_2) \in R$, it is written $r(c_1) = c_2$.
3. $A^o$ is a set of axioms on $O$.

In recent times, there is an increased use of ontologies in software engineering. The use of ontologies has been proposed by different researchers' in. the field of requirements engineering and management According to [40] the increased use can be attributed to the following: (i) they facilitate the semantic interoperability and (ii) they facilitate machine reasoning. Researchers have so far proposed many different synergies between software engineering and ontologies. In Requirements Engineering (RE), ontology can be used for: 1) describing requirements specification documents and 2) to formally represent requirements knowledge [10]. Ontology is an important resources of domain knowledge, especially in a specific application domain. In the management of IMRs, ontology can provides shared knowledge which can be useful in the management of IMRs in similar or cross domain knowledge management. By conceptualizing domain knowledge including the identified implicit requirement, it enables the easy adoption and identification and also management of IMRs. This reduces enormous costs in requirement development process.in making "explicit specification" it aids the reduction of ambiguous requirements and incomplete definitions during the elicitation process [40]. By using such ontology, several kinds of semantic processing can be achieved in requirements analysis [31]. In this work, ontology is considered a valid solution approach, because it has the potential to facilitate formalized semantic description of relevant domain knowledge for identification and management of IMR.

### C. Text Mining and Natural Language Processing

Text mining is the process of analyzing text to extract information that is useful for particular purposes [32]. [41] further expanded on the definition, stating that Text mining is the discovery and extraction of interesting, non-trivial knowledge from free or unstructured text everything from information retrieval (document or web site retrieval) to text classification and clustering, to entity, relation, and event extraction. It extracts information through the identification and exploration of interesting patterns [17]. Text mining has strong connections with Knowledge Management, data mining and Natural Language Processing (NLP). Authors in [41] describes NLP as an attempt to extract a fuller meaning representation from free text. In simple terms, it is figuring out who did what to whom, when, where, how and why. It 'typically makes use of linguistic concepts such as part-of-speech (noun, verb, adjective, etc.) and grammatical structure (either represented as phrases like noun phrase or prepositional phrase, or dependency relations like subject-of or object-of). NLP covers different disciplines from Linguistics to Computer Science and it's often closely linked with Artificial Intelligence. There are different definitions of NLP and they have evolved over the years. Natural Language Processing (NLP) generally refers to a range of theoretically motivated computational techniques for analyzing and representing naturally occurring texts [7].

According to [4] it is made up of the following sub areas which are linked to linguistics; i) Morphology ii) Syntax iii)Semantics iv)Pragmatics

The core purpose of NLP techniques is to realize human-like language processing for a range of tasks or applications [8]. The core NLP models used in this research are part-of-speech (POS) tagging and sentence parsers [7]. POS tagging involves marking up the words in a text as corresponding to a particular part of speech, based on both its definition, as well as its context. In addition, sentence parsers transform text into a data structure (called a parse tree), which provides insight into the grammatical structure and implied hierarchy of the input text [7].

NLP is used for our purpose in analysis of requirements statements to gain an understanding of similarities that exist between requirements and/or identify a potential basis for analogy. NLP in combination with ontology enables the extraction of useful knowledge from natural language requirements documents for the early identification and management of potential IMRs.

## III. RELATED WORK

Different researchers have proposed various ways for identification of IMRs. While some have presented applications, others have presented theoretical and conceptual frameworks and others take on the investigative approach in order to get real life views of practicing software engineers, requirements engineers and other specialists in the field on the practicality of stated theories, ideologies and concepts. Authors in [15] carried out a two part research aimed at identifying the impact of tacit and explicit knowledge transferred during software development projects. The first part involved an inductive, exploratory, qualitative methodology aimed at validating the tacit knowledge spectrum in software development projects. This involved unstructured interviews for data collection, and therefore assessment in a narrative form. The second part of this research involved the development of a conceptual framework of a model that supports future software development projects in their tacit to explicit knowledge transfers. [23] developed an approach based on a novel combination of grounded theory and incident fault trees. It focuses on Security Requirements. As a result of the threat landscape, there are new tacit knowledge which arise. This research proposes an approach to discover such unknown-knowns through multi-incident analysis. For this research an analysis and investigation method was used. It involved refining theoretical security models so that they remain pertinent in the face of a continuously changing threat landscape.

In a study carried out by [30], using a case study, the Knowledge Management on a Strategy for Requirements Engineering (KMoS-RE) which was designed to face the problem of management of tacit knowledge (in elicitation and discovery stage) and obtain a set of requirements that fulfill the clients' needs and expectations, was compared to requirements elicitation process proposed by MoProSoft; a Mexican software process model oriented to the specific needs of the software industry in Mexico. The results of this analysis showed that KMoS-RE seems to be more suitable than process proposed by MoProSoft. The KMoS-RE strategy improved the negotiation process and understanding about the domain and the software functionality requested and, the number of concepts and relationships was greater. KMos-RE strategy reduces the symmetry of ignorance between clients and users, and developers which facilitates the transference and transformation of knowledge and reduces increases the presence of unambiguous functional requirements.

Using requirements reuse for discovery and management of IMRs has been covered by a few studies. A study that draws on an analogy-making approach in managing IMRs is presented in [21]. This study proposes a system that uses semantic case-based reasoning for managing IMR. The model of a tool that facilitates the management of IMRs through analogy-based requirements reuse of previously known IMRs is presented. The system comprises two major components: semantic matching for requirements similarity and analogy-based reasoning for fine-grained cross domain reuse. This approach ensures the discovery, structured documentation, proper prioritization, and evolution of IMR, which is expected to improve the overall success of software development processes. However, as of now, this has not been adopted in a practical form. The work in [25] presents a model for computing similarities between requirements specifications to promote their analogical reuse. Hence, requirement reuse is based on the detection on analogies in specifications. This model is based on the assumption of semantic modeling abstractions including classification generalization and attribution. The semantics of these abstractions enable the employment of general criteria for detecting analogies between specifications without relying on other special knowledge. Different specification models are supported simultaneously. The similarity model which is relatively tolerant to incompleteness of specifications improves as the semantic content is enriched and copes well with large scale problems. Although the identification of analogies in requirements is essential, this study does not discuss the subject for the management of requirements. A method to highlight requirements potentially based on implicit or implicit-like knowledge is proposed in [2]. The identification is made possible by examining the origin of each requirement, effectively showing the source material that contributes to it. It is demonstrated that a semantic-level comparison enabling technique is appropriate for this purpose. The work helps to identify the source of explicit requirements based on tacit-like knowledge but it does not specifically categorize tacit requirement and its management. Also, in MaTREx [22], a brief review and interpretation of the literature on implicit knowledge useful for requirement engineering is presented. The authors describe a number of techniques that offer analysts the means to reason the effect of implicit knowledge and improve quality of requirements and their management. The focus of their work is on evolving tools and techniques to improve the management of requirements information through automatic trace recovery; discovering presence of tacit knowledge from tracking of presuppositions, non-provenance requirements etc. However, MaTREx still deals more with handling implicit knowledge.

Previous work on commonsense knowledge includes the Cyc project [9] with a goal to codify millions of pieces of common sense knowledge in machine readable form that enable a machine to perform human-like reasoning on such knowledge. Another source is WordNet [5] in which nouns and adjectives are manually organized into lexical classes, furthermore a distinction is made between words and word senses; yet its limitation is that there are no semantic relationships between the nouns and adjectives with the exception of extremely sparse attribute relations. Another prominent collection of commonsense is ConceptNet [20], which consists mainly of crowd sourced information. ConceptNet is the outcome of Open Mind Common Sense (OMCS) [6]. OMCS has distributed this CSK gathering task across general public on the Web. Through the OMCS website, volunteer contributors can enter CSK in natural language or even evaluate CSK entered by others.

Given this overview of the related literature, our proposed research stands out due to the fact that it introduces CSK for early identification and management of IMR. Moreover, it also embeds text mining and ontology to develop a support tool for managing IMRs. This is described next.

## IV. THE COTIR FRAMEWORK

The architectural framework is made up of three core technologies: text mining/NLP, CSK and ontology as presented in Fig. 1. The core system functionalities are depicted as rectangular boxes, while the logic, data and knowledge artefacts that enable core system functionalities are represented using oval boxes. The detailed description of COTIR is given in below.

### A. IMR Identification and Extraction

The part of the COTIR architecture that deals with knowledge representation and extraction is described in this section.

#### 1) Data Preprocessing

A Software Requirements Specification (SRS) document that has been preprocessed serves as input to the framework. Preprocessing is a manual procedure that which entails extraction of boundary sentences from the requirements document and further representing images, figures, tables etc. in its equivalent textual format for use by the system.

#### 2) NL Processor

The NL processor component facilitates the processing of natural language requirements for the process that enables feature extractor. The core natural language processing operations implemented in the architecture are i) Sentence selection, ii) Tokenization iii) Parts of speech (POS) tagging iv) Entity detection v) Parsing.
The Apache OpenNLP library [27] for natural language processing was used to implement all NLP operations.

#### 3) Ontology Library

The ontology library and CSKB both make up the knowledge model of our framework. The ontology library serves as a storehouse for the various domain ontologies (.owl/.rdf). The domain ontologies are those that have been developed for specific purpose or those of business rules. The ontology library is implemented using Java Protégé 4.1 ontology API.

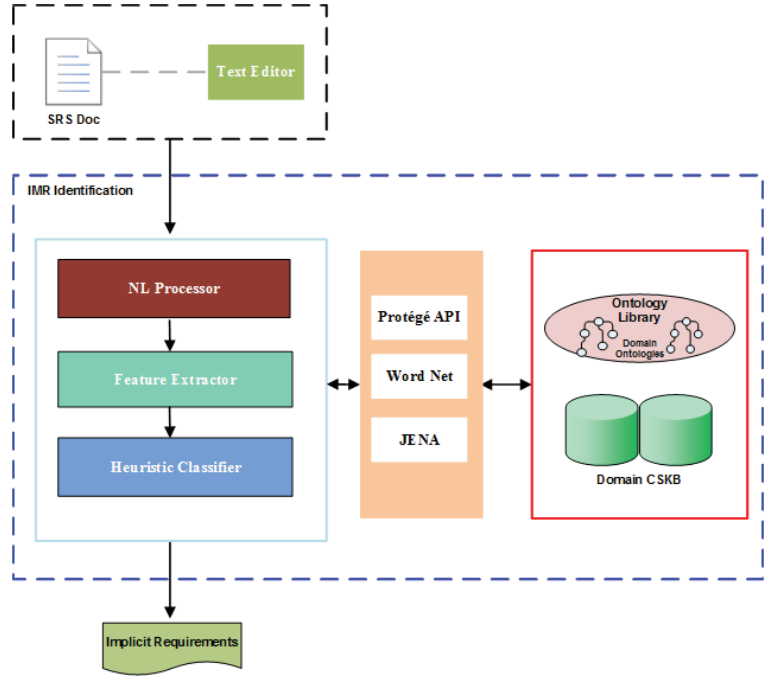

Fig. 1. Proposed COTIR Framework

#### 4) Common Sense Knowledge Base(CSKB)
The common sense knowledge bases of WebChild and domain-specific KBs are used for enhanced identification of IMR for specific domain.

#### 5) Feature Extractor
The feature extractor heuristic gives the underlying assumptions for identifying potential sources of IMR in a requirement document. Due to semantic features on which natural language text exist and by taking into account previous work done [11, 13, 16, 19, 28], the following characteristic features underline the significant aspects in a piece of text in terms of surface understanding that could possibly make a requirement implicit:
- Ambiguity such as structural and lexical ambiguity.
- Presence of vague words and phrases such as "to a great extent".
- Vague imprecise verbs such as "supported", "handled", "processed", or "rejected"
- Presence of weak phrases such as "normally", "generally".
- Incomplete knowledge.

## V. IMR SUPPORT TOOL USE AND EVALUATION

The COTIR framework illustrated in Fig. 1 is used to develop a support for the management of implicit requirements based on text mining, ontology and common sense knowledge. We now describe the use of this support tool for managing IMRs, followed by a snap shot of the tool in Fig. 2 and 3 then its performance evaluation.

### A. Use of the COTIR Tool

The process of using the COTIR tool developed in this work is as follows.

*Step 1:* Preprocess the source documents to get the requirements into text file format and devoid of graphics, images and tables.
*Step 2:* Select existing CSKB to be used for the identification of IMR.
*Step 3:* Import the requirement documents and domain ontology into COTIR environment.
*Step 4:* Click on the "analyze" function of the tool to allow the feature extractor identify potential sources of IMR in the requirement document.
*Step 5:* See the potential IMRs that are identified as well as their recommended possible explicit requirements.
*Step 6:* Seek the expert opinion on the IMRs; the experts could approve or disapprove the recommendations by adding/removing the recommendations through COTIR.

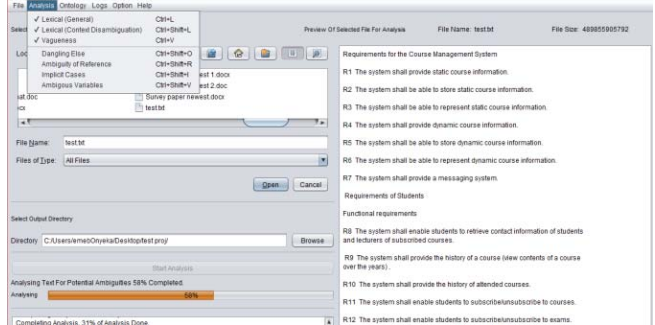

**Fig. 2. A Demo Snapshot of the COTIR Tool input/Analysis screen**

**Fig. 3. A Demo Snapshot of the COTIR Tool output**

For evaluation of the COTIR (Commonsense Ontology and Text-mining of Implicit Requirement) tool developed, we conduct an assessment of its performance using requirements specification. The objectives of the evaluation are as follows: (1) to assess the performance of the tool by human experts, (2) to determine its usefulness as a support tool for implicit requirements management within an organization. (3) to identify areas of possible improvement in the tool.

### B. Performance Evaluation Procedure

The evaluation makes use of the following sets of requirements specification: i) Course Management System (CMS) [33], this project was developed for use at the University of Twente course management system. The requirements for CMS describes basic functionality like course enrollment, course material and roster upload, student grading and e-mails communication. ii) EMbedded Monitoring Project [34], the EMMON project is a European Research and Development (R&D) project. The project captures requirements for smart locations and ambient intelligent environments (smart cities, smart homes, smart public spaces, smart forests, etc.) iii) Tactical Control System (TCS) requirements [35], This project was designed for the Naval Surface Warfare Center-Dahlgren Division and Joint Technology Center/System Integration Laboratory, Research Development and Engineering Center, U.S.

This three requirements specification documents were code named R1, R2, R3 as shown in the evaluation table II.

We use the following metrics to assess the performance of the system. Precision (P), Recall (R), F-measure (F).

$$R = \frac{TP}{(TP+FN)} \qquad P = \frac{TP}{TP+FP}$$

$$F = \frac{PR}{P + R}$$

In these formulas, TP, TN, FP and FN are as follows.
*TP (true positives):* number of requirements judged by both the expert and tool as being implicit
*TN (true negative):* number of requirements judged by both the expert and tool as not being implicit
*FP (false positive):* number of requirements judged by the tool as implicit and by the expert as not implicit
*FN (false negatives):* number of requirements judged by the tool as not implicit and by the expert as implicit.

A group of subjects were asked to mark implicitness in the requirement document and also use the tool.

The Subjects are a group of computing professionals, comprising software developers, academics and research students. They were given the following instructions: 1) for each specified requirement, mark each requirement based on its implicit nature (noting that a requirement may contain more than one form of implicitness). 2) For each requirement specify the degree of criticality of each implicitness on a scale of 1 to 5. (1 being least critical to 5 being most critical).

Table I shows a sample identification form. The type of implicitness includes i) Ambiguity (A) ii) Incomplete Knowledge (IK) iii) Vagueness (V) iv) Others

## C. Results of Performance Evaluation

Table II shows the recall, precision and F-scores computed for the tool relative to eight experts' (E1–E8) evaluation.

For a detection tool, the recall value is definitely more important than precision. In the ideal case, the recall should be 100%, as it would allow to relieve human analysts from the clerical part of document analysis [36]. For our tool with an average recall value of about 73.7%, it show that the tool is fit for practical use, as it marks six out of eight IMR detected by humans and is consistent with best practices. The average precision is 68.22% which shows the percentage of IMR judged by experts that was also retrieved by the tool is good and still consistent with best practices. The average F-score which is a harmonic mean of Precision and Recall is 70.3%. Which shows that the result of the tool's performance was good. As for the IMRs marked by human evaluators but missed by the tool, manual examination has shown that they represent implicit factors where we could not identify explicit patterns that would allow to automate IMR detection. Our observation from the simulation experiment (see Figs. 3.) is that the performance of the tool also depends significantly on the quality of the domain ontology and CSKB.

**Table I: Sample Identification Form**

| S/N | Requirement | Type of Implicitness | Criticality |
|---|---|---|---|
| 1 | The C&C shall provide the users with real--time data regarding the measured values, as collected from the various sensors part of the network. | (A)<br>(IK)<br>(V)<br>(O) | 1 2 3 4 5<br>1 2 3 4 5<br>1 2 3 4 5<br>1 2 3 4 5 |
| 2 | The C&C shall support the configuration of ranges for sensor readings (maximum and minimum allowed values). | (A)<br>(IK)<br>(V)<br>(O) | 1 2 3 4 5<br>1 2 3 4 5<br>1 2 3 4 5<br>1 2 3 4 5 |
| 3 | The C&C shall report potential sensor malfunctions to the users, when the reading is "Suspicious" or "Invalid". | (A)<br>(IK)<br>(V)<br>(O) | 1 2 3 4 5<br>1 2 3 4 5<br>1 2 3 4 5<br>1 2 3 4 5 |

**Table II: Recall, Precision and F-Score metrics from 8 experts (E1-E8)**

|  | Requirements | E1 | E2 | E3 | E4 | E5 | E6 | E7 | E8 | Average |
|---|---|---|---|---|---|---|---|---|---|---|
| Precision | R1 | 75 | 75 | 75 | 69.23 | 66.67 | 83.33 | 50 | 91.67 | 73.24 |
|  | R2 | 66.67 | 58.33 | 33.33 | 58.33 | 83.33 | 58.33 | 75 | 75 | 63.54 |
|  | R3 | 68.75 | 81.25 | 56.25 | 75 | 43.75 | 86.67 | 50 | 81.25 | 67.87 |
| Average |  |  |  |  |  |  |  |  |  | 68.22 |
| Recall | R1 | 90 | 90 | 75 | 90 | 80 | 83.33 | 75 | 78.57 | 82.74 |
|  | R2 | 66.67 | 70 | 66.67 | 58.33 | 76.92 | 70 | 69.23 | 75 | 69.1 |
|  | R3 | 73.33 | 72.22 | 64.29 | 66.67 | 58.33 | 81.25 | 61.54 | 76.47 | 69.26 |
| Average |  |  |  |  |  |  |  |  |  | 73.7 |
| F-Score | R1 | 81.82 | 81.82 | 75 | 78.26 | 72.73 | 83.33 | 60 | 84.62 | 77.2 |
|  | R2 | 66.67 | 63.63 | 44.44 | 58.33 | 80 | 63.63 | 72 | 75 | 65.46 |
|  | R3 | 70.97 | 76.47 | 60 | 70.59 | 50 | 83.87 | 55.17 | 78.79 | 68.23 |
| Average |  |  |  |  |  |  |  |  |  | 70.3 |

## VI. CONCLUSIONS

In conclusion, the ability to automatically identify and manage IMRs will mitigate many risks that can adversely affect system architecture design and project cost. This research work evolves a systematic tool support framework which uses common sense knowledge that can be integrated into an organizational Requirement Engineering procedure for identifying and managing IMR in systems development process. This is a direct response to problems in the practice of many organizations that have not been addressed by existing requirements management tools. Hence, this work addresses the problem of identifying IMRs in Requirements documents and its further management. The novelty of this work is that integrates three core technologies, namely, common sense knowledge, ontology and text mining to propose an automated IMR framework. Another significant contribution is that a support tool is developed based on the proposed framework and is helpful in domain-specific contexts. This work would useful to AI scientists and software engineers. Its targeted applications include providing software requirements for various AI systems, where common sense is useful in automation.

ACKNOWLEDGMENTS

This work was conducted when Onyeka Emebo was a Fullbright Scholar at Montclair State University, USA, visiting from Covenant University, Nigeria. The authors thank the source of the Scholarship for this funding.